\immediate\write18{makeindex -s nomencl.ist -o "\jobname.nls" "\jobname.nlo"}
\documentclass[a4paper,fleqn]{cas-dc}

\usepackage[sort&compress,numbers]{natbib}
\usepackage[nodots]{numcompress}

\usepackage{tikz}
\usetikzlibrary{calc}
\usepackage{graphicx}
\usepackage{subcaption}
\usepackage{amsmath}
\usepackage{tudcolours}

\newcommand{\rev}[1]{\ignorespaces #1}
\newcommand{\revB}[1]{\ignorespaces #1}
\newenvironment{revision}{\ignorespaces}{}
\usepackage{framed}
\usepackage{multicol} % Multiple columns environment
\usepackage{nomencl}
\makenomenclature
\usepackage{xstring}
\newcommand{\nomenclheader}[1]{%
  \item[\hspace*{-\itemindent}\normalfont\bfseries #1]}
\renewcommand\nomgroup[1]{%
  \IfStrEqCase{#1}{%
   {A}{\nomenclheader{Roman Symbols}}%
   {G}{\nomenclheader{Greek Symbols}}%
   {Z}{\nomenclheader{Subscripts}}%
  }%
}

\def\tsc#1{\csdef{#1}{\textsc{\lowercase{#1}}\xspace}}
\tsc{WGM}
\tsc{QE}
\tsc{EP}
\tsc{PMS}
\tsc{BEC}
\tsc{DE}
\usepackage{fancyhdr}
\usepackage{lastpage}
\fancypagestyle{plain}{\fancyhead[C]{}}
\begin{document}
\let\WriteBookmarks\relax
\def\floatpagepagefraction{1}
\def\textpagefraction{.001}

\shorttitle{Capacitance calculation of unloaded rolling elements}
\shortauthors{S. Puchtler et~al.}

\title [mode = title]{Capacitance calculation of unloaded rolling elements -- Comparison of analytical and numerical methods}

\author[1]{S. Puchtler}[orcid=0000-0002-2680-0140]
\cormark[1]
\ead{steffen.puchtler@tu-darmstadt.de}
\credit{Writing -- original draft, Methodology, Formal analysis, Visualisation}

\address[1]{Technische Universität Darmstadt, Institute for Product Development and Machine Elements, Otto-Berndt-Str. 2, 64287 Darmstadt, Germany}

\author[2]{T. Schirra}[orcid=0000-0001-8348-4327]
\ead{schirra@hcp-sense.com}
\credit{Investigation, Data Curation}

\address[2]{HCP Sense GmbH, Otto-Berndt-Str. 2, 64287 Darmstadt, Germany}

\author[1]{E. Kirchner}[orcid=0000-0002-7663-8073]
\ead{kirchner@pmd.tu-darmstadt.de}
\credit{Conceptualisation, Supervision}

\author[3,4]{Y. Späck-Leigsnering}[orcid=0000-0002-8302-802X]
\ead{spaeck@temf.tu-darmstadt.de}
\credit{Writing -- Review \& Editing}

\author[3,4]{H. De\ Gersem}[orcid=0000-0003-2709-2518]
\ead{degersem@temf.tu-darmstadt.de}
\credit{Software, Supervision}

\address[3]{Technische Universität Darmstadt, Institute for Accelerator Science and Electromagnetic Fields, Schloßgartenstr. 8, 64289 Darmstadt, Germany}

\address[4]{Technische Universität Darmstadt, Graduate School Computational Engineering, Dolivostr. 15,
64293 Darmstadt, Germany}

\cortext[cor1]{Corresponding author}

\begin{abstract}
Various methods for calculating the capacitance \rev{of unloaded rolling elements are compared to improve the electric characterization of rolling element bearings}. Semi-analytical approximations and finite element simulations are applied and a closed-form analytical two-dimensional solution is derived for comparison. The results show that the \rev{most common} semi-analytical approximation\rev{, which uses effective radii to express the contact geometry,}  reproduces the trend from numerical and analytical calculations, but \rev{overestimates the results derived from numerical simulations. Another semi-analytical approach is presented that delivers better results for the capacitance of unloaded point contacts.}
\end{abstract}

\begin{highlights}
\item Comparison of \rev{possible methods to calculate the capacitance of unloaded rolling elements:} semi-analytical and numerical methods
\item Derivation of an analytic 2D formula for calculating the capacitance per length unit of an unloaded ball-raceway contact
\item Comparison of multiple interpretations of the semi-analytical capacitance calculation
\item \rev{Including the rim area aside of the groove into the capacitance calculation}
\end{highlights}


\maketitle

\section{Introduction}
The electrical properties of rolling bearings are increasingly important: With an increasing number of \rev{variable frequency drives}, electrically induced bearing damage rises. Therefore, research in electrical bearing failures is a ``key bottleneck in electric vehicles at present and in the forthcoming decades'' \cite{He.2020}. The electrical characterization of bearings \rev{helps to} predict and prevent \rev{current induced} bearing damages in electric machines \cite{Mutze.2016}. \rev{Until today, multiple} studies introduced electric models of rolling bearings \cite{Busse.1997,Mutze.2004,Mutze.2007,Gemeinder.2014, Furtmann.2017, Graf.2020}. Furthermore, electrical bearing characteristics can be used sensorially \rev{for load estimation. The electrical properties} depend on the lubricant film thickness and, thus, on the load condition \cite{Schirra.2018, Schirra.2019}, which in turn requires a precise electrical modeling of the bearing \rev{and allows the utilization of rolling element bearings as sensor integrating machine elements \cite{Kirchner.2018}. However, every model uncertainty contributes to measurement uncertainty of the sensorially utilized bearing. Thus, improved} electrical models \rev{have been introduced} for sensorial applications \cite{Schnabel.2016,Maruyama.2019,Schirra.2021}. In addition, capacitance measurements are used for lubrication film investigations as they have certain advantages compared to tactile or optical measurements \cite{Zhang.2021}. \rev{Therefore, precise} electrical models \rev{are key} to improve capacitive film thickness measurements \cite{Glovnea.2012,Cen.2014,Jabonka.2018}.

The work at hand contributes to a better understanding of \rev{\revB{different capacitance calculation methods with the aid of the unloaded rolling element geometry. This provides a basic understanding that can be applied to loaded elements as well. Therefore,} commonly used semi-analytic equations} \revB{are compared} with a newly derived analytic formula \rev{for a two dimensional (2D) setting, and with a three dimensional (3D)} finite element (FE) simulation. In addition, different implementations of the semi-analytic formula are compared and discussed.
	
\subsection{Bearing capacitance}
\label{seq:bearingCapacitance}
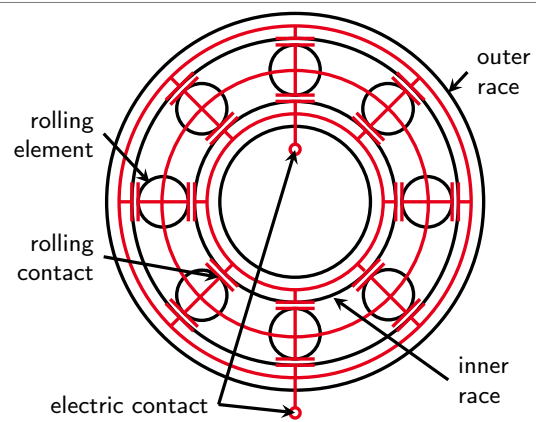
\begin{figure}
	\centering
	\begin{tikzpicture}
		\node [above right, inner sep=0] (image) at (0,0) {\includegraphics[width=.6\linewidth]{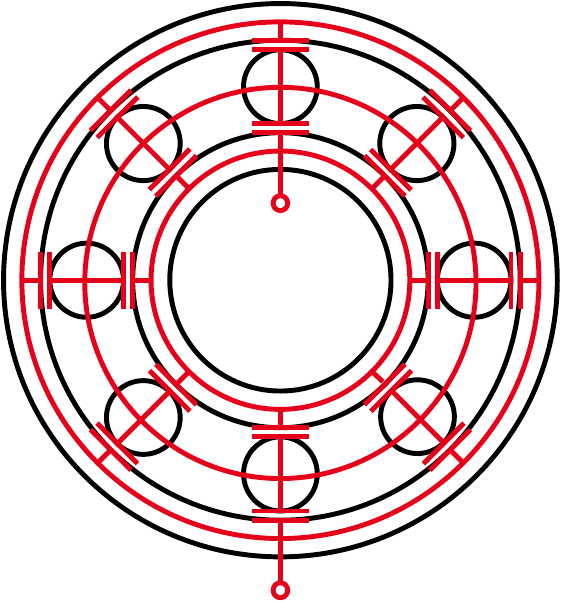}};
		\begin{scope}[
			x={($0.1*(image.south east)$)},
			y={($0.1*(image.north west)$)}]
			%\draw[lightgray,step=1] (image.south west) grid (image.north east);

			\draw[stealth-, very thick] (6,3) -- (9,1) node[right,black,align=left,fill=white]{inner \\ race};
			\draw[stealth-, very thick] (9,8) -- (9.5,8.5) node[right,black,align=left,fill=white]{outer \\ race};
			\draw[stealth-, very thick] (5,6.6) -- (3,0.4) node[left,black,fill=white]{electric contact};
			\draw[stealth-, very thick] (5,0.2) -- (3,0.4);
			\draw[stealth-, very thick] (3,3.5) -- (0,4) node[left,black,align=right,fill=white]{rolling \\ contact};
			\draw[stealth-, very thick] (1.2,5.8) -- (0,7) node[left,black,align=right,fill=white]{rolling \\ element};
		\end{scope}
	\end{tikzpicture}
	\caption{\rev{Simplified electrical model of a rolling bearing, assuming single contact capacitances for every rolling element-race contact and a conductive cage \revB{\cite{Schirra.2018}}.}}
	\label{fig:CapacityNetwork}
\end{figure}

\rev{Considering a} radially loaded deep groove ball bearing in fluid friction regime \rev{with a} high resistivity \rev{lubricant}, the dominant electric characteristic for frequencies in and beyond the $\mathrm{kHz}$ regime is the permittivity $\varepsilon$. Fig. \ref{fig:CapacityNetwork} shows an example of a capacitor network of a deep groove ball bearing with a conductive cage based on the model of Prashad \cite{Prashad.1988}. Radially loaded bearings develop a load zone where rolling elements carry the load. \rev{This work studies the capacitance of unloaded rolling elements out of the load zone.}

\rev{The geometry of unloaded rolling elements is comparable to the area outside the contact ellipse of loaded rolling contacts.} Different authors, i.e. \cite{Maruyama.2019,Schirra.2021,Prashad.1988,Jabonka.2012,Schneider.2021,Prashad.2006,Magdun.2009,Leenders.1987}, used a semi-analytical method (\ref{eq:CapacitanceSemianalytic}) to compute the capacitance of these areas,
	\begin{equation}
		C_{\mathrm{semi}} = \iint_A \frac{\varepsilon}{h(x,y)}  \, \mathrm{d}x\mathrm{d}y .
		\label{eq:CapacitanceSemianalytic}
	\end{equation}
    \nomenclature[Z]{semi}{Semi-analytical}%
\begin{revision}%
Thereby, the electrode area of the capacitor, $A$, is decomposed in infinitesimal area elements $\mathrm{d}A$ that are typically expressed in Cartesian coordinates as $\mathrm{d}A=\mathrm{d}x\mathrm{d}y$, where $x$ and $y$ act as a parametrization of $A$.
The varying distance between the electrodes $h(x,y)$ is expressed by a Taylor expansion of second order in terms of the minimal distance between the electrodes $s$ and the effective radii $R_x$ and $R_y$,
    \begin{equation}
        h(x,y) = s + \frac{x^2}{2R_x} + \frac{y^2}{2R_y} .
        \label{eq:h_xy}
    \end{equation}
In each direction, the effective radius is calculated according to Hertz theory \cite{Hertz.1882,Hamrock.1981} from the radii of the two contact partners $r_1$ and $r_2$,
	\begin{equation}
		R = \frac{1}{\frac{1}{r_1} + \frac{1}{r_2}} .
		\label{eq:EquivalentRadius}
	\end{equation}
    %\nomenclature[Z]{semi}{Semi-analytical}%

\subsection{Scope and assumptions}
\label{sec:scopeAssumptions}

This work investigates the following assumptions of the semi-analytic approach (\ref{eq:CapacitanceSemianalytic})-(\ref{eq:EquivalentRadius}):

\begin{itemize}
    \item The equivalent geometry derived from the Hertzian effective radii results in an equal capacitance.
    \item The error of using the second-order Taylor expansion instead of the true geometry is small.
    \item The electric field lines are staight and perpendicular to $A$ and thus, the description as a parallel connection of infinitesimal capacitances $\mathrm{d}C=\varepsilon \mathrm{d}a / h$ is legitimate.
    \item The edge of the raceway groove has no significant influence on the total capacitance.
\end{itemize}

\end{revision}

For this purpose, different calculation models are implemented in addition to \rev{the semi-analytic} methods, i.e., a 2D analytical model as well as a 2D and a 3D finite element (FE) model. These are compared with each other and the deviations of the calculated capacitances are quantified.

\rev{In the following, we assume a constant} electric permittivity $\varepsilon$ resulting in a non-coupled and linear problem. The ball-raceway contact is assumed to be completely \rev{immersed in lubricant as in \cite{Schirra.2021,Magdun.2009}}. Approaches to account for the oil distribution around the contact can be found in \cite{Jabonka.2012,Schneider.2021}. \revB{The centrifugal forces on the unloaded rolling elements are neglected as they range below $7~\mathrm{N}$ for the considered 6205 ball bearing at a rotational speed of up to $10,000~\mathrm{rpm}$. Consequently, a sufficient lubrication gap between the rolling element and the outer ring is assumed, separating both contact partners.}
Furthermore, the contact surfaces are assumed to be ideally smooth as is usually the case in literature \cite{Mutze.2007,Magdun.2009,Schirra.2021,Schneider.2021}. \rev{Effects attributed to surface roughness are considered beyond the scope of this paper.} Insights on rough rolling surfaces are given in \cite{Gemeinder.2014,Schnabel.2016}.
	
\section{Theory and Calculation}

\begin{revision}
\subsection{Bearing geometry}
 Table \ref{tab:GeometryParameters} shows the parameters of the 6205 bearing geometry and  the corresponding values used in this manuscript. %
\begin{table}[width=.95\linewidth,cols=2]
	\caption{\rev{Geometry parameters of a 6205 deep groove rolling bearing used for calculations.}}
	\label{tab:GeometryParameters}
	\begin{tabular*}{\tblwidth}{@{} CC@{} }
		\toprule
		\multicolumn{2}{C}{\textbf{Section Plane I}} \\
        \hline
		Inner ring & Outer ring\\
        \hline
		\multicolumn{2}{C}{$r_{1y}=r_\mathrm{re}=4~\mathrm{mm}$} \\
        \hline
		$r_{2y,\mathrm{i}}=-r_\mathrm{g,i}=-4.16~\mathrm{mm}$ & $r_{2y,\mathrm{o}}=-r_\mathrm{g,o}=-4.24~\mathrm{mm}$ \\
		\hline
        $R_{y,i}=104~\mathrm{mm}$ & $R_{y,o}=70.67~\mathrm{mm}$ \\
        \hline
        $g_\mathrm{g,i}=5.03~\mathrm{mm}$ & $g_\mathrm{g,o}=4.82~\mathrm{mm}$ \\
        \hline
        \multicolumn{2}{C}{$B=15~\mathrm{mm}$} \\
		\toprule
		
        \multicolumn{2}{C}{\textbf{Section Plane II}} \\
        \hline
        Inner ring & Outer ring \\
        \hline
        \multicolumn{2}{C}{$r_{1x}=r_\mathrm{re}=4~\mathrm{mm}$} \\
        \hline
        $r_{2x,\mathrm{i}}=r_\mathrm{rw,i}=15.25~\mathrm{mm}$ & $r_{2x,\mathrm{o}}=-r_\mathrm{rw,o}=-23.25~\mathrm{mm}$ \\
        \hline
        $R_{x,i}=3.17~\mathrm{mm}$ & $R_{x,o}=4.83~\mathrm{mm}$ \\
        \toprule
        \multicolumn{2}{C}{\textbf{Geometry}} \\
        \hline \\
        \multicolumn{2}{C}{
        	\centering
        	\begin{tikzpicture}
        		\node [above right, inner sep=0] (image) at (0,0) {\includegraphics[width=.95\linewidth]{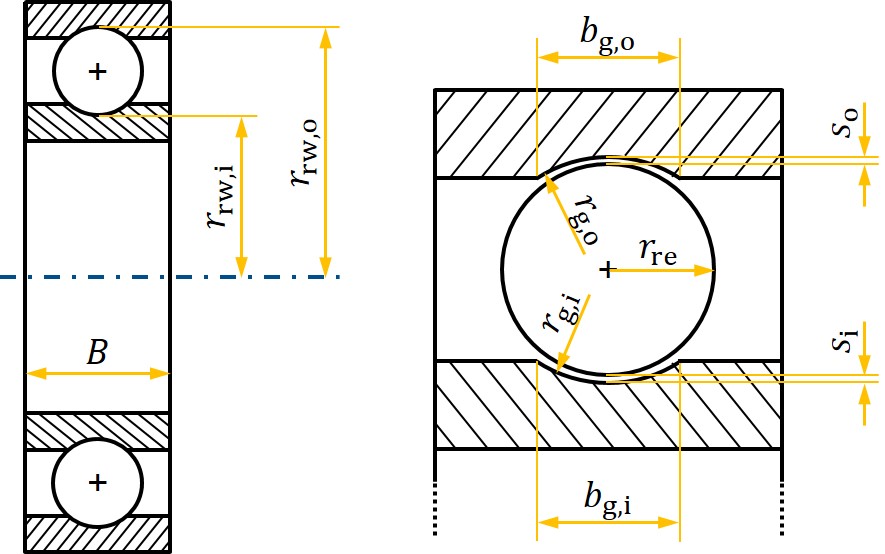}};
        		\begin{scope}[
        			x={($0.1*(image.south east)$)},
        			y={($0.1*(image.north west)$)}]
        			%\draw[lightgray,step=1] (image.south west) grid (image.north east);
        		\end{scope}
        	\end{tikzpicture}
        } \\
		
		\bottomrule
	\end{tabular*}%
\end{table}%
\end{revision}%
\rev{In the 2D setting, all effects scale with the geometric size. Hence, a simplification of the formulae can be achieved by }dividing all geometry parameters by the rolling element's radius $r_{\mathrm{re}}$. For easier recognition, all \rev{scaled} parameters are noted in Greek letters, Table \ref{tab:DimensionlessGeometry}. \rev{Two sectional planes are distinguished in this work: section plane I, corresponding to the ($\eta$,$\zeta$) plane and section plane II, corresponding to the ($\xi$,$\zeta$) plane.} The minimal distance between race and rolling element $s$ is described by the dimensionless quantity $\alpha=s/r_{\mathrm{re}}$. $\beta=b_\mathrm{g}/r_{\mathrm{re}}$ expresses the groove width, and $\tau$ the groove radius. \rev{The groove radius} is derived from the race conformity $f$, which in return is the ratio between groove radius and ball diameter \rev{in section plane I. In section plane II, $\tau$ is derived from the} raceway radius $r_\mathrm{rw}$. $\sigma$ is the distance between race and ball center and is calculated from the quantities above, as shown in Table \ref{tab:DimensionlessGeometry}, which shows the orientation of the dimensionless coordinate system ($\xi$,$\eta$,$\zeta$) as well. In a polar coordinate system, $\varrho=r/r_{\mathrm{re}}$ describes the dimensionless radial coordinate.
In both section planes, the rolling element and raceway represent non-concentric circular sections as electrodes. For these, a capacitance per length unit is to be calculated. As this is scalable, one parameter can be eliminated effectively by representing the geometry without dimensions.
Furthermore, the problem is generalized for the inner and outer ring by assuming a negative radius for the raceway radius on the inner ring $r_\mathrm{r,i}$, from which negative dimensionless parameters $\sigma$ and $\tau$ result.

	\begin{table*}[width=.9\textwidth,cols=3]
		\caption{Dimensionless contact geometry with raceway radius $\tau$, distance between center points of rolling element and raceway radius $\sigma$, minimal distance between rolling element and race $\alpha$ and race width $\beta$.}
		\label{tab:DimensionlessGeometry}
		\begin{tabular*}{\tblwidth}{@{} CCC@{} }
			\toprule
			\multirow{2}{*}{Section plane I} & \multicolumn{2}{c}{Section plane II}\\		
			\cline{2-3}
			& Inner race & Outer race\\
			\midrule
				\centering
				\begin{tikzpicture}
					\node [above right, inner sep=0] (image) at (0,0) {\includegraphics[width=.2\textwidth]{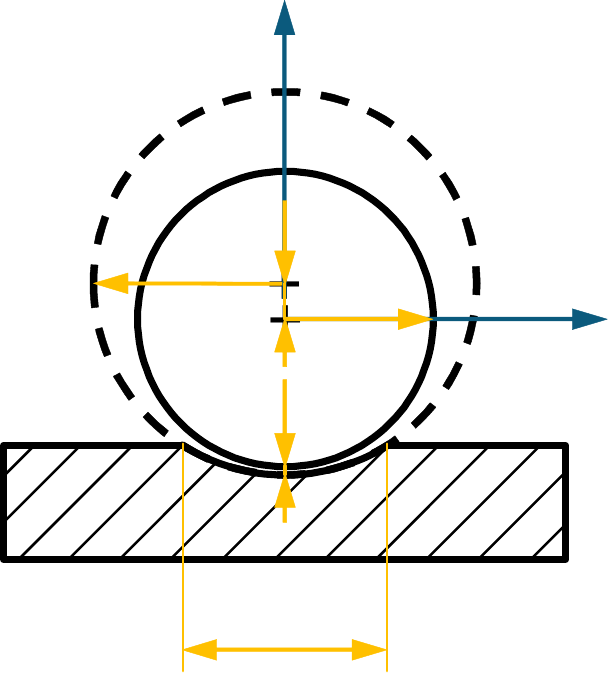}};
					\begin{scope}[
						x={($0.1*(image.south east)$)},
						y={($0.1*(image.north west)$)}]
						%\draw[lightgray,step=1] (image.south west) grid (image.north east);
		
						\draw (9.5,5.7) node{$\eta$};
						\draw (5.2,9.5) node{$\zeta$};
						\draw (5.9,5.7) node{$\rev{\varrho =} 1$};
						\draw (3.5,6.2) node{$\tau$};
						\draw (4.2,6.5) node{$\sigma$};
						\draw (4.2,3.9) node{$\alpha$};
						\draw (4.8,0.8) node{$\beta$};
					\end{scope}
				\end{tikzpicture}
				&
				\begin{tikzpicture}
					\node [above right, inner sep=0] (image) at (0,0) {\includegraphics[width=.22\textwidth]{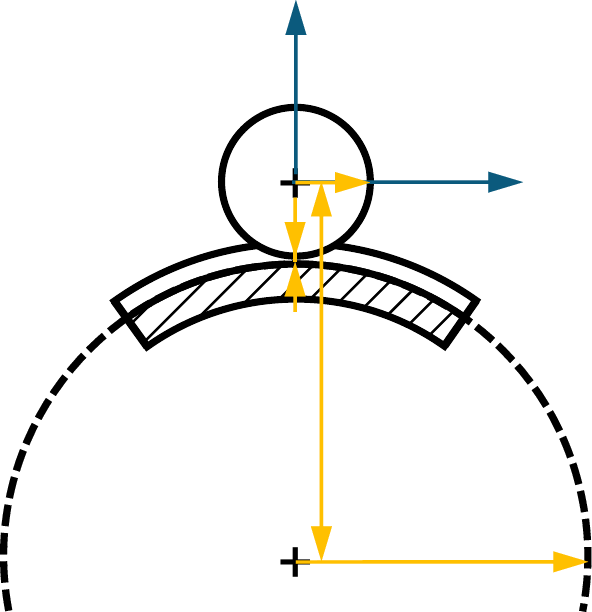}};
					\begin{scope}[
						x={($0.1*(image.south east)$)},
						y={($0.1*(image.north west)$)}]
						%\draw[lightgray,step=1] (image.south west) grid (image.north east);
		
						\draw (8.5,7.6) node{$\xi$};
						\draw (5.5,9.5) node{$\zeta$};
						\draw (5.5,7.5) node{$1$};
						\draw (7.5,1.3) node{|$\tau$|};
						\draw (5,4) node{|$\sigma$|};
						\draw (4.6,6.5) node{$\alpha$};
					\end{scope}
				\end{tikzpicture}
				&
				\begin{tikzpicture}
					\node [above right, inner sep=0] (image) at (0,0) {\includegraphics[width=.33\textwidth]{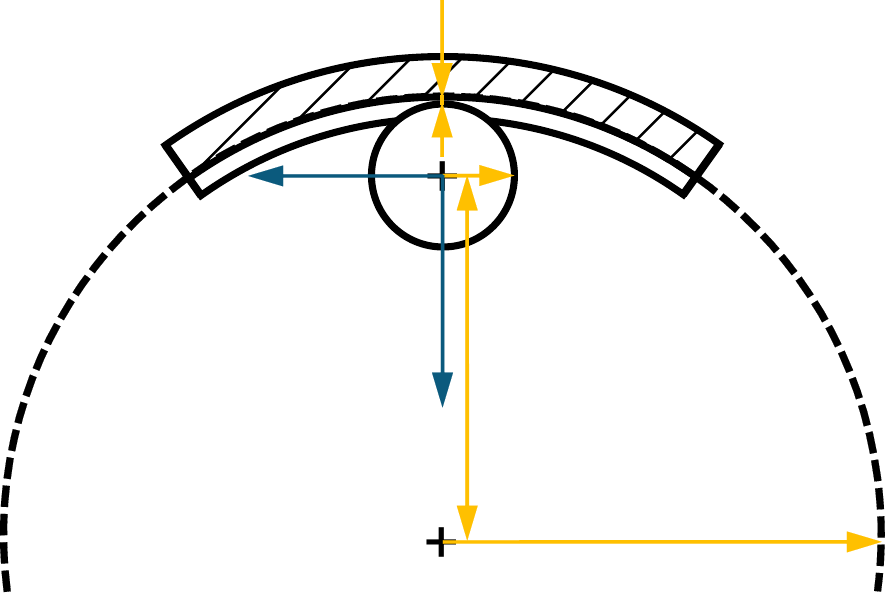}};
					\begin{scope}[
						x={($0.1*(image.south east)$)},
						y={($0.1*(image.north west)$)}]
						%\draw[lightgray,step=1] (image.south west) grid (image.north east);
		
						\draw (3.2,6.3) node{$\xi$};
						\draw (4.7,3.7) node{$\zeta$};
						\draw (5.4,7.5) node{$1$};
						\draw (7.5,1.3) node{$\tau$};
						\draw (5.6,4) node{$\sigma$};
						\draw (4.7,9.5) node{$\alpha$};
					\end{scope}
				\end{tikzpicture}\\

			\midrule
				$\tau = -r_\mathrm{g}/r_\mathrm{re} > 0$
				&
				$\tau = -r_{\mathrm{rw,i}}/r_{\mathrm{re}} < 0$
				&
				$\tau = -r_{\mathrm{rw,o}}/r_{\mathrm{re}}  > 0$ \\
				$\sigma = \tau - 1 - \alpha > 0$
				&
				$\sigma = \tau - 1 - \alpha < 0$
				&
				$\sigma = \tau - 1 - \alpha > 0$ \\
				$\beta = b_{\mathrm{g}}/r_{\mathrm{re}}$ \\
			\bottomrule
		\end{tabular*}
	\end{table*}

\subsection{\rev{Geometry mapping of the common semi-analytic approximation}}
\begin{revision}
The capacitances of regions outside the Hertzian area of loaded rolling elements are often calculated by the semi-analytical approximation (\ref{eq:CapacitanceSemianalytic})-(\ref{eq:EquivalentRadius}), in the following referred to as model A\label{model:A}. A summary of models introduced in this work is listed in Appendix \ref{seq:appendix}.
Model A uses the effective radii derived from the Hertzian theory \cite{Hertz.1882,Hamrock.1981} in both dimensions opposing a flat plain. For later comparison, a closed and exact solution in 2D of of the per length unit capacitance of this configuration is used,
    \begin{equation}
        C'_\mathrm{eff} = \frac{2\pi\varepsilon}{\ln{\left(\frac{R}{R+s-\sqrt{(2R+s)s}}\right)}} .
        \label{eq:capPlaneCyl}
    \end{equation}
This can be interpreted as a circle of \rev{the radius $R$, (\ref{eq:EquivalentRadius}),} in a distance $s$ from an infinite plane. The following equation is used to represent the true geometry derived from the per length capacitance of an eccentric cylinder capacitor \cite{Chen.2009},
    \begin{equation}
        C'_\mathrm{true} = \frac{2\pi \varepsilon}{\ln{\left(\frac{r_\mathrm{2}^2-r_\mathrm{1}^2+e^2-\sqrt{(r_\mathrm{2}^2-r_\mathrm{1}^2+e^2)^2-4r_\mathrm{2}^2e^2}}{r_\mathrm{2}^2-r_\mathrm{1}^2-e^2-\sqrt{(r_\mathrm{2}^2-r_\mathrm{1}^2-e^2)^2-4r_\mathrm{1}^2e^2}}\frac{r_\mathrm{1}}{r_\mathrm{2}}\right)}} ,
        \label{eq:capExcCyl}
    \end{equation}
where $r_1$ and $r_2$ are the radii of the electrodes and $e$ is the distance between the center points of both cylindrical electrodes, which can be expressed as $e=s-r_\mathrm{2}-r_\mathrm{1}$.

The semi-analytic solution of the equivalent radius configuration can be obtained by (\ref{eq:CapacitanceSemianalytic}) and
    \begin{equation}
        \begin{matrix}
            h(x) = s + R_x - \sqrt{R_x^2-x^2} \\
            h(y) = s + R_y - \sqrt{R_y^2-y^2}
        \end{matrix} \: .
        \label{eq:h_realR_x}
    \end{equation}
This relation is often simplified by using a second-order Taylor series approximation \cite{Prashad.2006}
    \begin{equation}
        \begin{matrix}
            h(x) = s + x^2/2R_x \\
            h(y) = s + y^2/2R_y
        \end{matrix} \: .
        \label{eq:h_TaylorR_x}
    \end{equation}
    
\end{revision}

\subsection{Alternative realizations of the semi-analytic approximation}
\label{seq:semiMethods}
 \rev{For the semi-analytic approximation, (\ref{eq:CapacitanceSemianalytic}), the contact region} is divided into plate capacitors of an infinitesimally small area $\mathrm{d}A=\mathrm{d}x\mathrm{d}y$ with variable height $h(x,y)$ and then integrated. However, this equation can be interpreted in various ways. For example, the arrangement of the infinitesimal plate capacitors, as well as the integration limits, can be chosen in different manners.

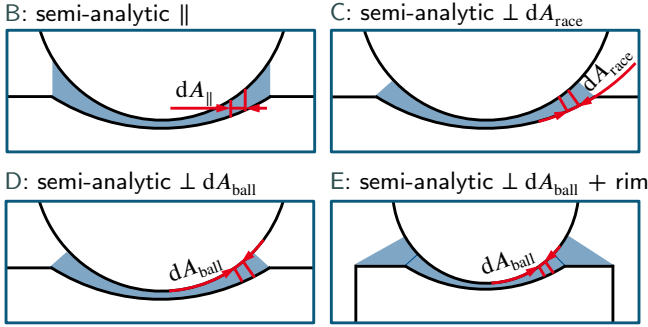
\begin{figure}[pos=h]
		\centering
		\begin{tikzpicture}
			\node [above right, inner sep=0] (image) at (0,0) {\includegraphics[width=\linewidth]{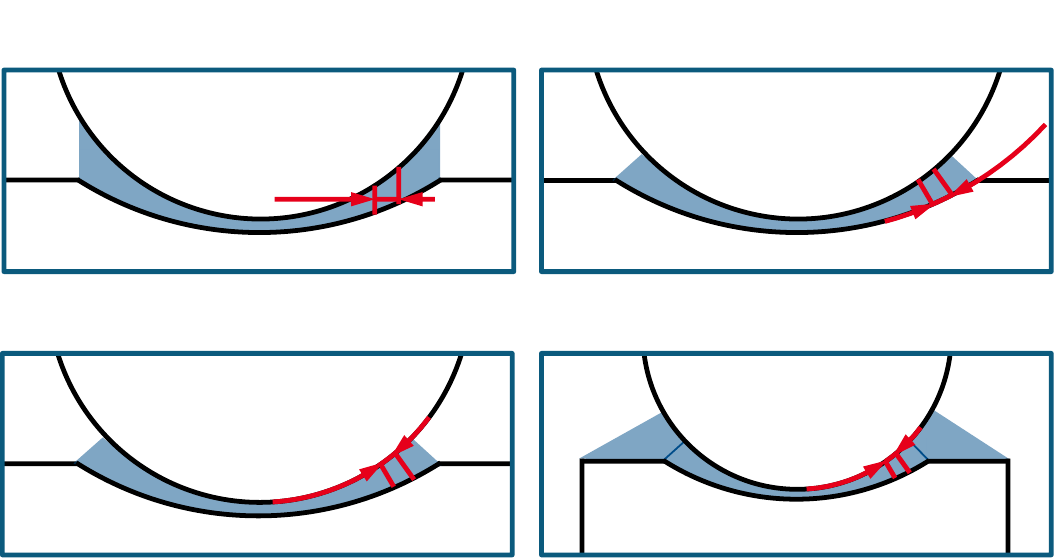}};
			\begin{scope}[
				x={($0.1*(image.south east)$)},
				y={($0.1*(image.north west)$)}]
				%\draw[lightgray,step=1] (image.south west) grid (image.north east);
				
				\draw (3,6.88) node{$\mathrm{d}A_{\|}$};
				\draw (9.45,7.6) node[rotate=45]{$\mathrm{d}A_{\mathrm{race}}$};
				\draw (3,1.7) node[rotate=20]{$\mathrm{d}A_{\mathrm{ball}}$};
				\draw (7.9,1.9) node[rotate=20]{$\mathrm{d}A_{\mathrm{ball}}$};
				
				\draw (0  ,9.2) node[anchor=west,inner sep=0pt]{\small \hyperref[model:B]{B}: semi-analytic $\|$};
				\draw (5.1,9.2) node[anchor=west,inner sep=0pt]{\small \hyperref[model:C]{C}: semi-analytic $\perp\mathrm{d}A_{\mathrm{race}}$};
				\draw (0  ,4.2) node[anchor=west,inner sep=0pt]{\small \hyperref[model:D]{D}: semi-analytic $\perp\mathrm{d}A_{\mathrm{ball}}$};
				\draw (5.1,4.2) node[anchor=west,inner sep=0pt]{\small \hyperref[model:E]{E}: semi-analytic $\perp\mathrm{d}A_{\mathrm{ball}}$ + rim};
			\end{scope}
		\end{tikzpicture}
		\caption{Different realizations of semi-analytic capacitance approximation in section plane I: Infinitesimal plate capacitors in parallel (model \hyperref[model:B]{B}) or perpendicular to the rolling element surface (models \hyperref[model:C]{C}-\hyperref[model:E]{E}), using area elements from the raceway surface ($\mathrm{d}A_{\mathrm{race}}$, model \hyperref[model:C]{C}) or from the rolling element surface ($\mathrm{d}A_{\mathrm{ball}}$, models \hyperref[model:D]{D} and \hyperref[model:E]{E}). Model \hyperref[model:E]{E} includes the rim area aside the groove.}
		\label{fig:DifferentSemianalyticMethods}
	\end{figure}

\rev{In this paper, three additional realizations of the 2D approximation according to (\ref{eq:CapacitanceSemianalytic}) are compared.} The method $\mathrm{d}A_{\|}$ (model B\label{model:B}), Fig. \ref{fig:DifferentSemianalyticMethods} left, uses mutually parallel infinitesimal plate capacitors \rev{on the true geometry without a Taylor series expansion.} The integration limits are chosen to be the groove width in section plane I and the rolling element diameter in section plane II.

For the \rev{following two methods}, Fig. \ref{fig:DifferentSemianalyticMethods} right, the height $h$ of the infinitesimal plate capacitors is chosen perpendicular to the rolling element surface. Method  $\perp\mathrm{d}A_{\mathrm{ball}}$ (model C\label{model:C}) then uses the surface element of the raceway for the capacitance calculation, method $\perp\mathrm{d}A_{\mathrm{race}}$ (model D\label{model:D}) on the other hand that of the rolling element. Again, the groove width is used as the integration limit in section plane I, see Table \ref{tab:DimensionlessGeometry}. In section plane II, the inner ring is integrated until the height $h(x,y)$ meets the inner ring tangentially, whereas $\pi/2$ is selected as the integration limit for the outer ring. \rev{Model \hyperref[model:D]{D} can be enhanced to take the rim area aside of the groove into account. Therefore, an additional capacitance contribution between the integration limits in section plane I of the groove edge and the end of the ring is calculated (Model E\label{model:E}).}

\subsection{2D analytic calculation}
\label{seq:2Dcalculus}
Within both relevant section planes \rev{of an unloaded ball-raceways contact, Table \ref{tab:DimensionlessGeometry}, both surfaces} conform to a pair of eccentric circle sections. For this arrangement an analytical solution (model F\label{model:F}) can be derived, following Chen \cite{Chen.2009}.

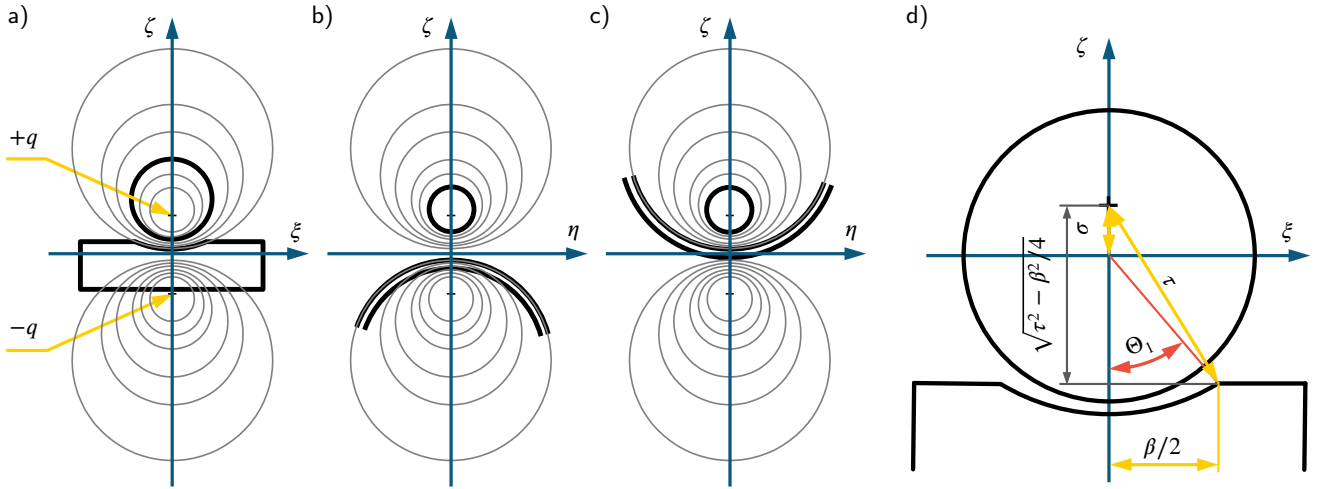
\begin{figure*}
		\centering
		\begin{tikzpicture}
			\node [above right, inner sep=0] (image) at (0,0) {\includegraphics[width=\linewidth]{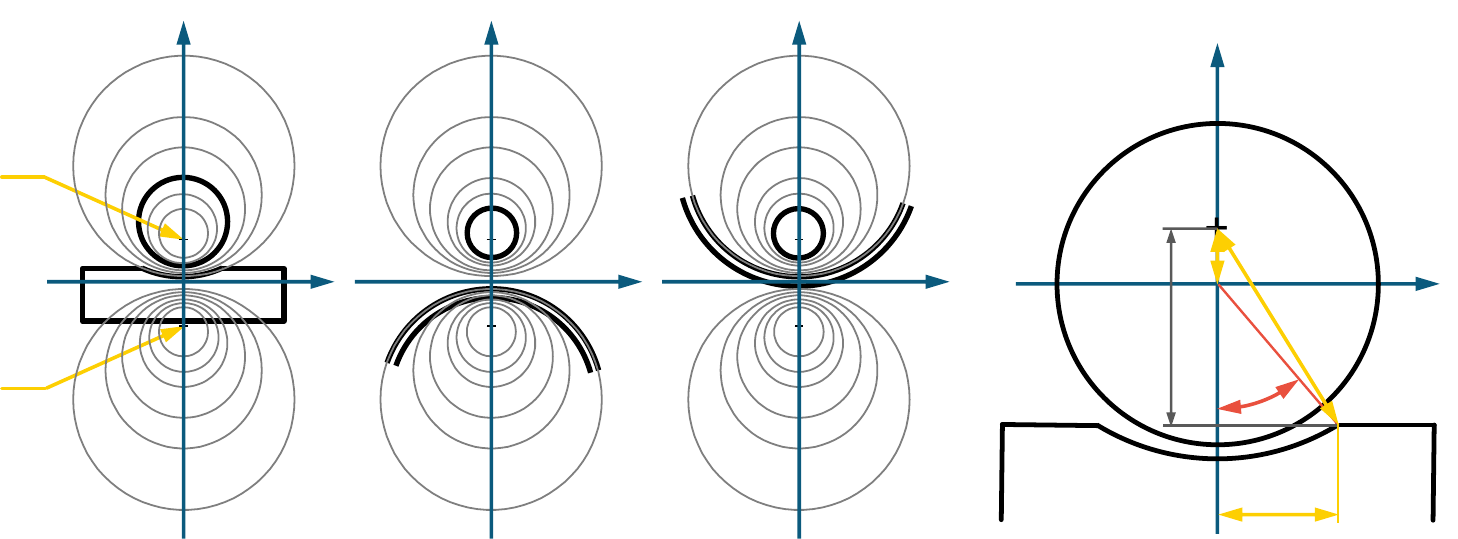}};
			\begin{scope}[
				x={($0.1*(image.south east)$)},
				y={($0.1*(image.north west)$)}]
				%\draw[lightgray,step=1] (image.south west) grid (image.north east);
				
				\draw (0.1,9.6) node{a)};
				\draw (2.4,9.6) node{b)};
				\draw (4.5,9.6) node{c)};
				\draw (6.9,9.6) node{d)};
				
				\draw (0.14,7.1) node{$+q$};
				\draw (0.14,3.2) node{$-q$};
				\draw (2.2,5.2) node{$\xi$};
				\draw (1.1,9.4) node{$\zeta$};
				\draw (4.3,5.2) node{$\eta$};
				\draw (3.2,9.4) node{$\zeta$};
				\draw (6.4,5.2) node{$\eta$};
				\draw (5.3,9.4) node{$\zeta$};
				
				\draw (9.7,5.2) node{$\xi$};
				\draw (8.15,9) node{$\zeta$};
				\draw (8.75,0.8) node{$\beta/2$};
				\draw (8.58,2.9) node[rotate=25]{$\Theta_1$};
				\draw (8.15,5.3) node[rotate=90]{$\sigma$};
				\draw (8.8,4.2) node[rotate=-60]{$\tau$};
				\draw (7.8,3.9) node[rotate=90]{$\sqrt{\tau^2-\beta^2/4}$};
			\end{scope}
		\end{tikzpicture}
		\caption{Equipotential lines of the potential field of two line charges with indicated electrode contours a) in section plane I and in section plane II b) at the inner ring c) at the outer ring. d) Integration limit $\Theta_1$ in section plane I.}
		\label{fig:AnalyticDerivition}
	\end{figure*}

First, an electric potential field must be found that fulfills the boundary conditions of the arrangement. The surfaces of the rolling element and the raceway act as electrodes and are assumed to be ideally conductive. A potential field that fulfills these boundary conditions can be achieved with two imaginary line charges with a charge coating of $+q$ and $-q$ respectively, see Fig. \ref{fig:AnalyticDerivition}.a-c. The resulting equipotential circles are also known as Apollonian circles \cite{Chen.2009}.

By varying the distance between the line charges and selecting two circles with matching radii, two equipotential circles can be determined which coincide with the electrode surfaces. To describe the relationship between both circles, the auxiliary quantity $\kappa$ is defined as
	\begin{equation}
		\kappa = \frac{1}{2\sigma}\left(\tau^2-\sigma^2-1-\sqrt{(\tau^2-\sigma^2-1)^2-4\sigma^2}\right) .
		\label{eq:kappa}
	\end{equation}
Then, the potential field can be determined in dimensionless polar coordinates as
	\begin{equation}
		\Phi(\varrho,\Theta)=\frac{q}{4\pi\varepsilon} \: \ln\left(\frac{\varrho^2-2\varrho\cos\Theta/\kappa + 1/\kappa^2}{\varrho^2-2\varrho\cos\Theta\cdot\kappa + \kappa^2} \right) .
		\label{eq:potentialField}
	\end{equation}

By forming gradients on the surface of the sphere ($\varrho=1$), the electric field strength $\left. E_\varrho \right|_{\varrho=1} = \left. -\frac{\partial\Phi}{\partial\varrho}\right|_{\varrho=1}$ can be calculated, which only has a radial component perpendicular to the electrode surface. Using the material law, the electric flux $\left. D_\varrho \right|_{\varrho=1} = \varepsilon \left. E_\varrho \right|_{\varrho=1}$ and finally, using Gauss' theorem, the charge per length unit $Q'$ on the rolling element can be determined,
	\begin{equation}
		Q'=\frac{q}{\pi}\int_0^{\Theta_1} \left. D_\varrho \right|_{\varrho=1} \mathrm{d}\Theta=\frac{q}{\pi}\arctan\left[\frac{1+\kappa}{1-\kappa}\tan\left(\frac{\Theta_1}{2}\right)\right] .
		\label{eq:charge}
	\end{equation}
	
The angle $\Theta_1$ is introduced as the integration limit in section plane I, which is calculated from the groove width $\beta$ according to Fig. \ref{fig:AnalyticDerivition}.d,
	\begin{equation}
		\Theta_1=\arctan\left(\frac{\beta/2}{\sqrt{\tau^2-\beta^2/4}-\sigma}\right) .
		\label{eq:ThetaLimit}
	\end{equation}
	
In section plane II, $\pi/2$ is \rev{chosen as} the integration limit so that only the side of the rolling element facing the ring is included.
Finally, the potentials of the two equipotential circuits are determined,
	\begin{align}
		\Phi_{\mathrm{ball}} &= \frac{q}{2\pi\varepsilon}\ln\left|\frac{1}{\kappa}\right| \text{and} \\
		\Phi_{\mathrm{race}} &= \frac{q}{2\pi\varepsilon}\ln\left|\frac{\tau-\sigma-1/\kappa}{\tau-\sigma-\kappa}\right| .
	\end{align}
Now the capacitance of the arrangement can be determined,
	\begin{equation}
		C'=\frac{Q'}{\Phi_{\mathrm{ball}}-\Phi_{\mathrm{race}}} = 4\varepsilon\:\frac{\arctan\left[\frac{1+\kappa}{1-\kappa}\tan\left(\Theta_1/2\right)\right]}{\ln\left|\frac{\tau-\sigma-\kappa}{(\tau-\sigma)\kappa-1}\right|} .
		\label{eq:AnalyticCapacity}
	\end{equation}

\begin{figure}[pos=h]
		\centering
		\begin{tikzpicture}
			\node [above right, inner sep=0] (image) at (0,0) {\includegraphics[width=\linewidth]{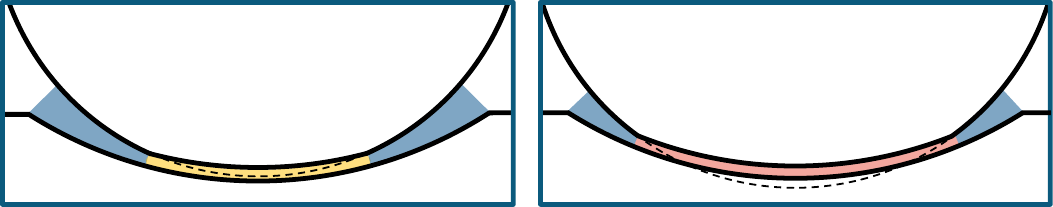}};
			\begin{scope}[
				x={($0.1*(image.south east)$)},
				y={($0.1*(image.north west)$)}]
				%\draw[lightgray,step=1] (image.south west) grid (image.north east);
			\end{scope}
		\end{tikzpicture}
		\caption{Visualization of acceptable (left) and critical deflection (right) for the 2D calculation.}
		\label{fig:2DcalculusLimitation}
	\end{figure}

Beside the calculation of the mentioned section planes in radial deep groove ball bearings, this equation can also be used for undeflected geometries of cylindrical roller bearings, needle roller bearings, or plain bearings. An extension by integrating the capacitance in the third spatial direction is also conceivable to represent unloaded rolling elements in tapered or spherical roller bearings.
In \rev{theory, it is possible to utilize the approach} to calculate the capacitance of edge areas of loaded rolling elements \rev{by introducing} a lower integration limit $\Theta_0$ for the end of the Hertzian surface can be specified in (\ref{eq:charge}). The resulting capacitance loading can alternatively be calculated with $C'_{\mathrm{res}}=C' (\Theta_1 )-C' (\Theta_0 )$. However, the deflection of the rolling element raceway contact must not be greater than the lubricant film thickness, because this would lead to overlapping equipotential circles in the calculation approach (see Fig. \ref{fig:2DcalculusLimitation}), which is a physical contradiction and consequently causes a mathematical contradiction. \rev{As in reality the deflection is usually greater than the lubrication film, this approach is not feasible for the edge area of loaded rolling elements.}

	\begin{figure}[pos=h]
		\centering
		\begin{tikzpicture}
			\node [above right, inner sep=0] (image) at (0,0) {\includegraphics[width=\linewidth]{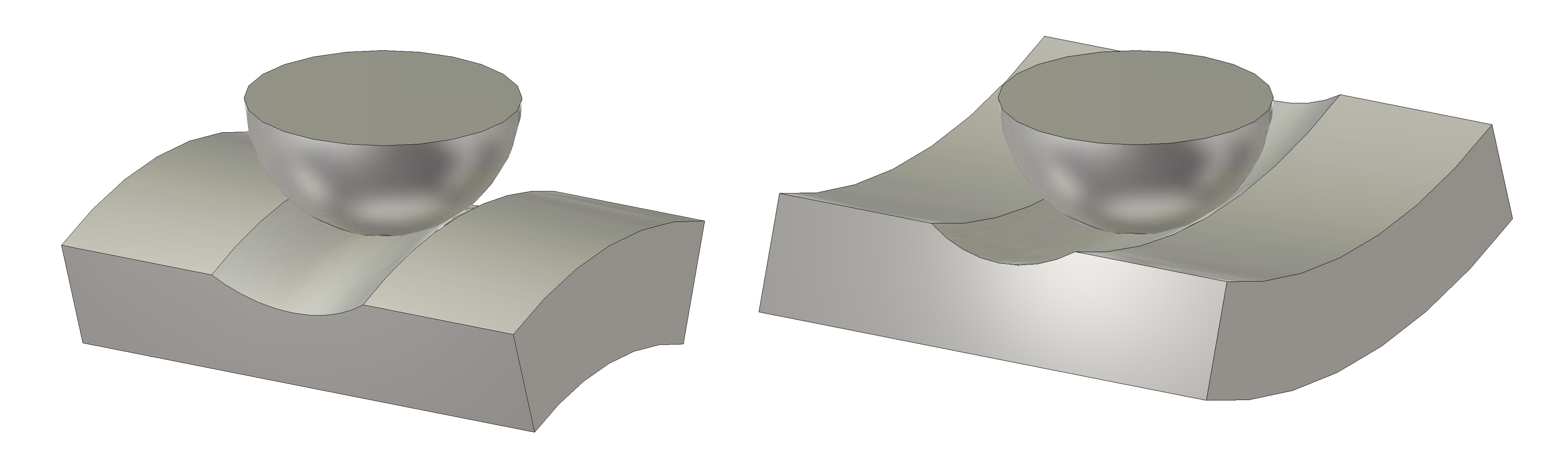}};
			\begin{scope}[
				x={($0.1*(image.south east)$)},
				y={($0.1*(image.north west)$)}]
				%\draw[lightgray,step=1] (image.south west) grid (image.north east);
			\end{scope}
		\end{tikzpicture}
		\caption{Geometries of the inner (left) and outer (right) race models used in the FE simulations.}
		\label{fig:NumericModels}
	\end{figure}

\subsection{Geometry and boundary conditions of implemented FE models}
\label{seq:FE_Model}
\begin{figure}
		\centering
		\begin{tikzpicture}
			\node [above right, inner sep=0] (image) at (0,0) {\includegraphics[width=\linewidth]{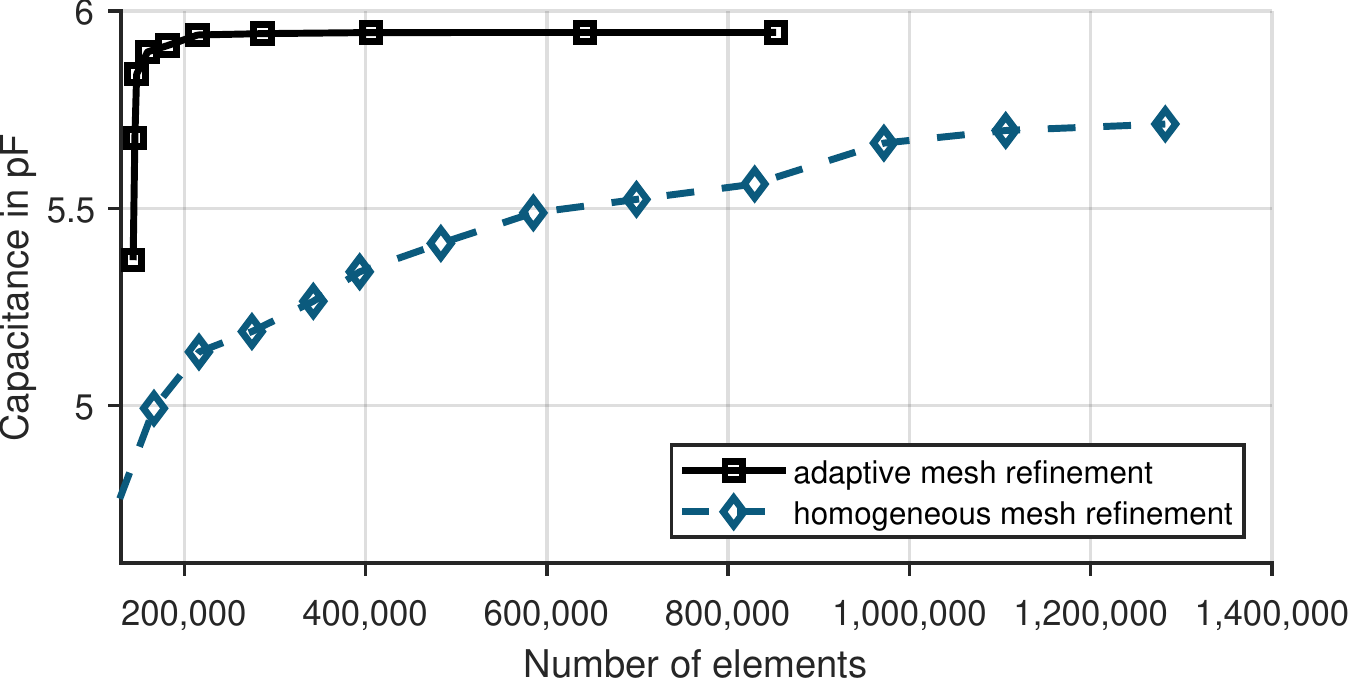}};
			\begin{scope}[
				x={($0.1*(image.south east)$)},
				y={($0.1*(image.north west)$)}]
			%\draw[lightgray,step=1] (image.south west) grid (image.north east);
			\end{scope}
		\end{tikzpicture}
		\caption{\rev{A convergence study using homogeneous and adaptive mesh refinement.}}
		\label{fig:meshstudy}
	\end{figure}

For complex geometries, an analytic solution can usually not be found. Instead, such geometries are \rev{approximated using semi-analytic approaches or} modeled and simulated numerically. In this work, a single contact between a half ball and a raceway cut \rev{has been simulated in 2D with an in-house FE solver (model G\label{model:G}) for comparison with the analytical model. 
Moreover, the single contact configuration has been simulated in 3D with the simulation tool CST Studio\textsuperscript{\copyright} (model H\label{model:H}) in order to dispose of a reference value for the true 3D geometry.}
The geometry parameters have been chosen from a 6205-C3 radial deep groove ball bearing on the inner and outer ring, \rev{as collected in Table \ref{tab:GeometryParameters} and Fig. \ref{fig:NumericModels}. 
The model has been reduced to one quarter by introducing both symmetry planes in order to reduce the computation time.} 
The bearing ring is \rev{grounded} and a potential of 1 V is applied to the rolling element section. The boundary face in direction of the shaft or housing is grounded with the corresponding bearing ring. 
All other edge faces of the model are assumed electrically insulating, which means that all electrical flux leaving the rolling element flows to the bearing ring. The electrostatic solvers calculate the electric field and from this the charges $+Q$ and $-Q$ on the electrodes. The capacitance then follows as $C=Q/(1\mathrm{V})$.
\rev{A convergence study was carried out in order to find an accurate and at the same time affordable FE simulation, Fig. \ref{fig:meshstudy}. Homogeneous mesh refinement converged slowly. The adaptive mesh refinement available in CST Studio\textsuperscript{\copyright} allowed to reach an accuracy better that $0.1\%$ with around $300,000$ tetrahedral elements. The mesh adaption started from a maximum mesh width of $262.5~\mu m$. The mesh was mostly refined in the center of the contact and at the edges of the groove.}

\section{Results and Discussion}
\begin{revision}
\subsection{Mapping of the groove geometry}
The most commonly used model \hyperref[model:A]{A} uses the effective radii $R_x$ and $R_y$ derived from the Hertzian theory instead of the true geometry to calculate the outside area capacitance. To estimate the possible error of this simplification, the difference of the closed analytical solution of the effective radius configuration, (\ref{eq:capPlaneCyl}), is compared to the one of the true geometry, (\ref{eq:capExcCyl}), in Fig. \ref{fig:realVsEqui}. To visualize the difference, the relative deviation of the capacitance is plotted as a function of the lubrication gap. Four cases are considered: section plane I ($R_y$) and section plane II ($R_x$) for both, the inner and the outer contact. For all cases, the deviation converges to zero for small lubrication gaps. Convex-convex pairs underestimate the true capacitance and the deviation increases with an increasing effective radius $R$. All in all, the deviations are below $1\%$ for lubrication gaps up to $5~\mathrm{\mu m}$ and therefore, the assumption to use the effective geometry for capacitance calculation is legitimate.

\begin{figure}[pos=tb]
	\centering
	\begin{tikzpicture}
		\node [above right, inner sep=0] (image) at (0,0) {\includegraphics[width=\linewidth]{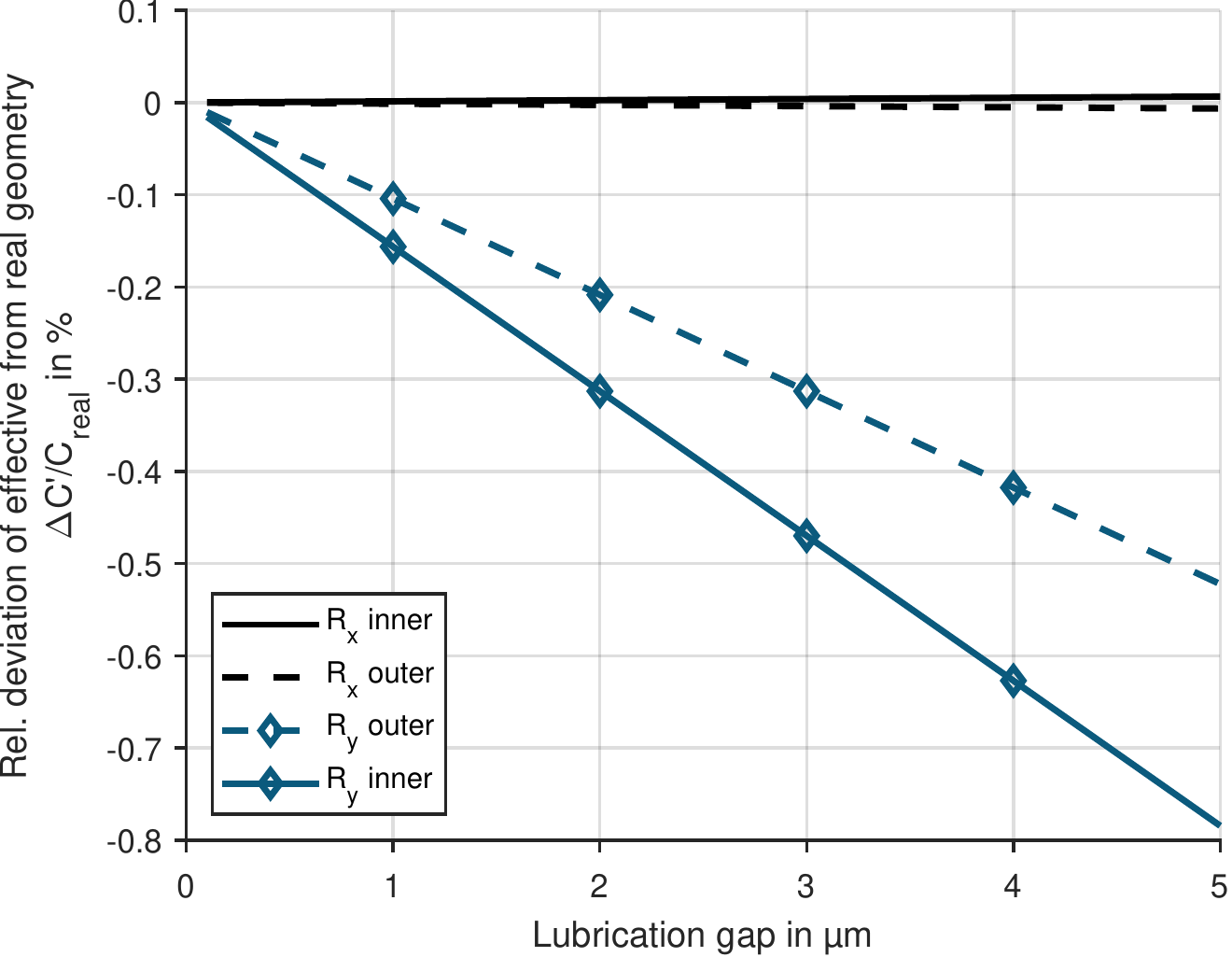}};
		\begin{scope}[
			x={($0.1*(image.south east)$)},
			y={($0.1*(image.north west)$)}]
			%\draw[lightgray,step=1] (image.south west) grid (image.north east);
		\end{scope}
	\end{tikzpicture}
	\caption{\rev{Relative deviation of the capacitance per length unit calculated with effective radii, (\ref{eq:capPlaneCyl}), from the solution of the true geometry, (\ref{eq:capExcCyl}), for section plane I ($R_y$) and section plane II ($R_x$) and the inner and outer raceway contact.}}
	\label{fig:realVsEqui}
\end{figure}

Another assumption is that the second-order Taylor series expansion approximates the height of the effective radius with a sufficient precision. In Fig. \ref{fig:squaredVsCircle}, the relative deviation of the second Taylor polynomial, (\ref{eq:h_TaylorR_x}), from the exact solution, (\ref{eq:h_realR_x}), is plotted for the effective radius configuration (left) and the true geometry (right). In both cases, the deviation converges to zero for small lubrication gaps and the Taylor polynomial overestimates the exact formulation. Thus, the errors of the two assumptions discussed so far partially cancel each other. For the effective geometry, the error caused by the Taylor approximation is less than $1\%$ for lubrication gap heights of less than $5~\mathrm{\mu m}$.
For the true geometry, the Taylor expansion is depreciated, as it results in a considerable error.

\begin{figure}[pos=tb]
	\centering
	\begin{tikzpicture}
		\node [above right, inner sep=0] (image) at (0,0) {\includegraphics[width=\linewidth]{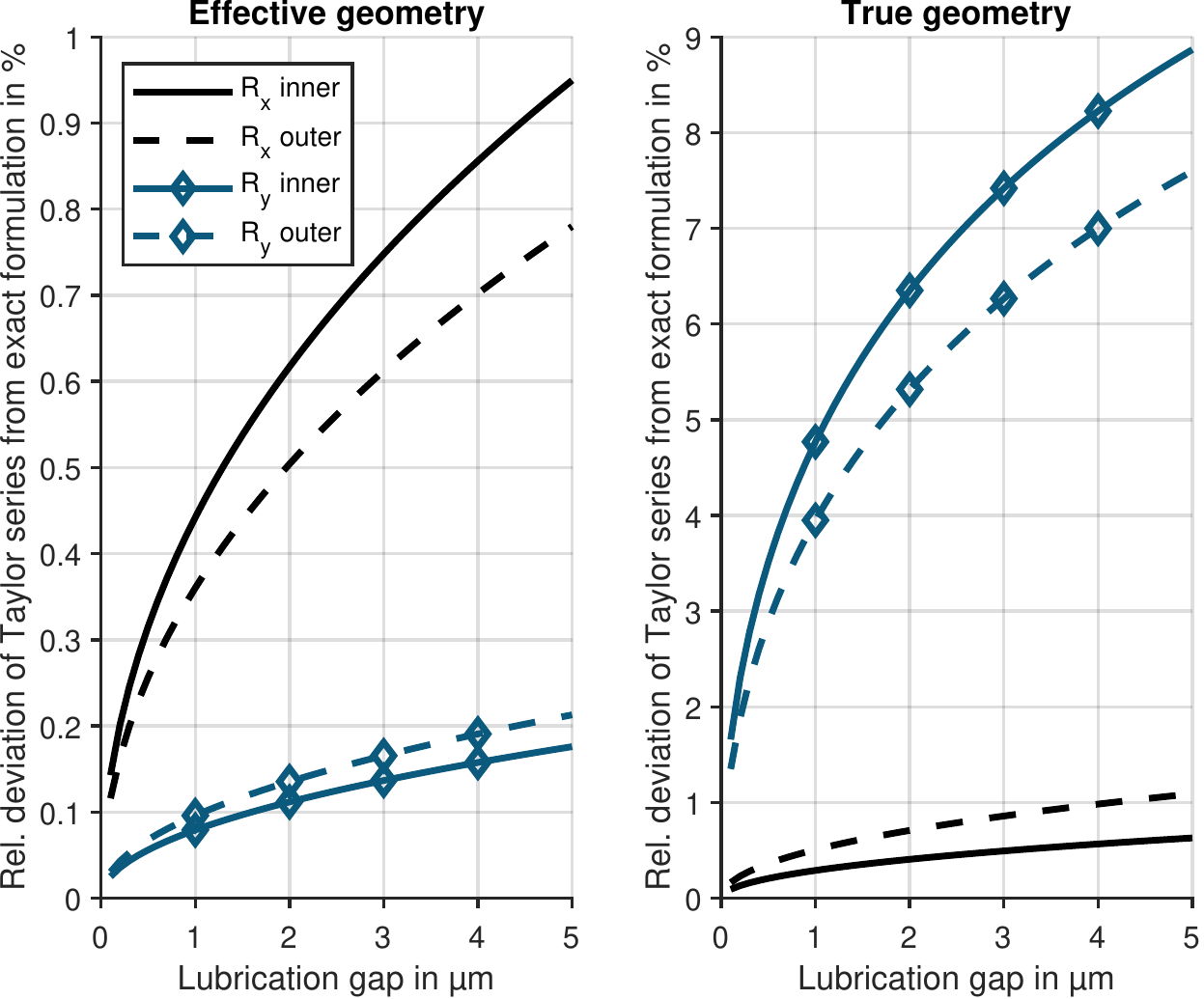}};
		\begin{scope}[
			x={($0.1*(image.south east)$)},
			y={($0.1*(image.north west)$)}]
			%\draw[lightgray,step=1] (image.south west) grid (image.north east);
		\end{scope}
	\end{tikzpicture}
	\caption{\rev{Relative deviation of the capacitance per length unit calculated with the semi-analytic approximation, (\ref{eq:CapacitanceSemianalytic}), with the hight $h(x)$ calculated using the second Taylor polynomial, (\ref{eq:h_TaylorR_x}), from the solution using the exact height, (\ref{eq:h_realR_x}), for section plane I ($R_y$) and section plane II ($R_x$) and the inner and outer raceway contact, each for the effective geometry (left) and the true geometry (right).}}
	\label{fig:squaredVsCircle}
\end{figure}

\end{revision}

\subsection{Comparison of 2D results}
\begin{figure*}
		\centering
		\begin{tikzpicture}
			\node [above right, inner sep=0] (image) at (0,0) {\includegraphics[width=\linewidth]{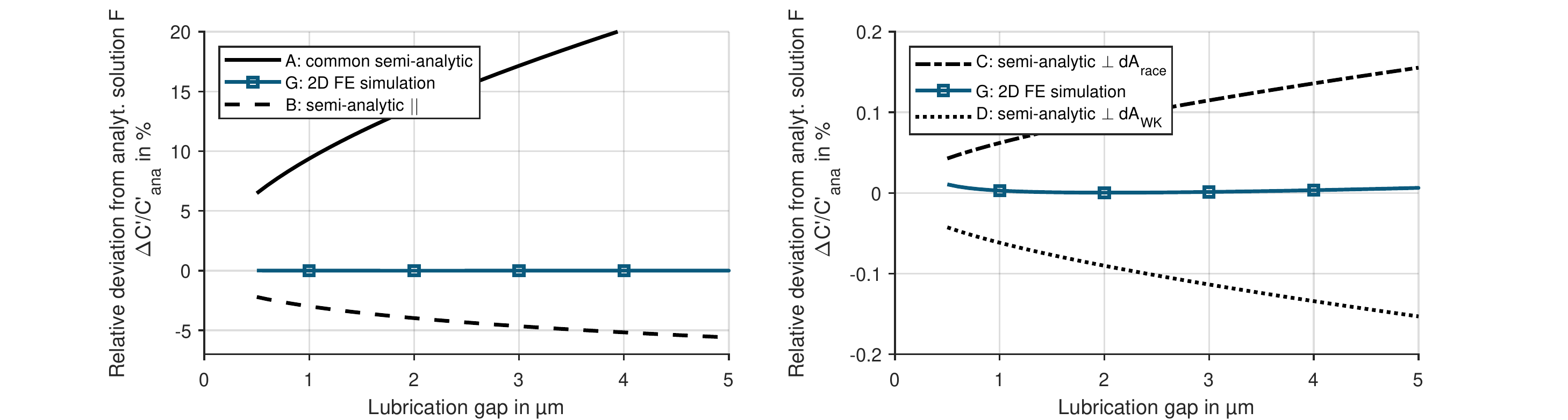}};
			\begin{scope}[
				x={($0.1*(image.south east)$)},
				y={($0.1*(image.north west)$)}]
				%\draw[lightgray,step=1] (image.south west) grid (image.north east);
			\end{scope}
		\end{tikzpicture}
		\caption{\rev{Relative deviation of different semi-analytic models \hyperref[model:A]{A}-\hyperref[model:D]{D} and 2D FE model \hyperref[model:G]{G}, from the analytic model \hyperref[model:F]{F} for the outer race of a 6205-C3 deep-groove ball bearing in section plane I, Table \ref{tab:DimensionlessGeometry}.}}
		\label{fig:2Dcomparison}
	\end{figure*}

\rev{Next, the results of the 2D calculations, the analytical model \hyperref[model:F]{F} according to Section \ref{seq:2Dcalculus} and the 2D FE model \hyperref[model:G]{G}, Section \ref{seq:FE_Model}, are compared to various realizations of the semi-analytic models \hyperref[model:A]{A}, \hyperref[model:B]{B}, \hyperref[model:C]{C} and \hyperref[model:D]{D}, Section \ref{seq:semiMethods}, in Fig. \ref{fig:2Dcomparison}. The relative deviation of the semi-analytical methods and the 2D FE model from the analytical solution is shown for the capacitance per length unit of bearing 6205-C3.} In the left diagram, it can be seen that \rev{the common semi-analytic method \hyperref[model:A]{A} overestimates the analytic solution by up to $10\%$ for lubrication gaps less than $1~\mathrm{\mu m}$. This is because it ignores the groove width as if the groove would the surround the whole ball, resulting in a too high capacitance. Method \hyperref[model:B]{B}}, semi-analytic $\|$, underestimates the capacitance. This is due to the fact that only the capacitor area projected in the $\zeta$ direction is taken into account and that with increasing distance from the center axis the height $h$ is no longer plotted at the shortest distance between the two electrodes. 

The right diagram in Fig. \ref{fig:2Dcomparison} shows an enlarged section of the left diagram. It is clear that the semi-analytic methods \hyperref[model:C]{C} ($\perp\mathrm{d}A_{\mathrm{race}}$) and \hyperref[model:D]{D} ($\perp\mathrm{d}A_{\mathrm{ball}}$) approximate the analytically calculated results better. When using the rolling element surface, the result is underestimated and when using the raceway surface, the result is overestimated, as the latter has the larger area. 
\rev{The remaining deviations of the 2D FE model \hyperref[model:G]{G} with respect to the analytical model \hyperref[model:F]{F} are attributed to the different shapes of the insulating boundary at the groove edge side. The analytic model truncates the domain at an electric field line of the eccentric circles solution, whereas the 2D FE model truncates the geometry at a straight line through the groove edge. However, the deviation in the considered range is small with about $0.01\%$.}
	
Yet, the results of the 2D calculations are not directly transferable to the real ball-raceway contact. The 2D methods assume two eccentric cylinder surfaces for the electrodes while the true geometry consists of a sphere for the ball and a torus cut-out for the groove. Since an analytical solution for this problem is not found, the semi-analytical \rev{methods \hyperref[model:A]{A} and \hyperref[model:E]{E} are} compared with a FE simulation in Section \ref{seq:comparison3D}.

\subsection{Influence of the rim area aside of the groove}
\begin{figure}[pos=tb]
		\centering
		\begin{tikzpicture}
			\node [above right, inner sep=0] (image) at (0,0) {\includegraphics[width=\linewidth]{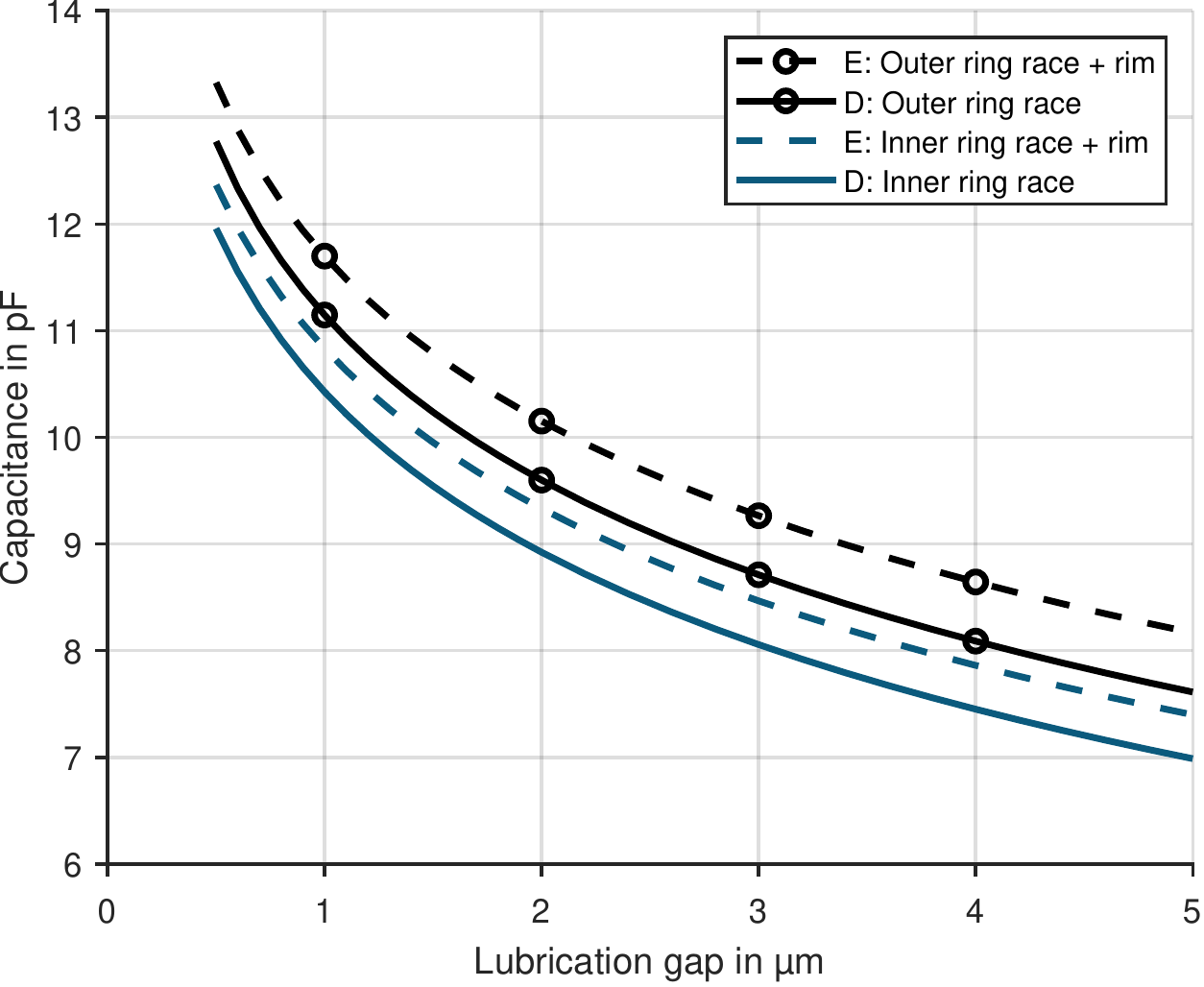}};
			\begin{scope}
				
			\end{scope}
		\end{tikzpicture}
		\caption{Comparison of capacitance over lubrication gap for \rev{the semi-analytic $\perp\mathrm{d}A_{\mathrm{ball}}$ models \hyperref[model:D]{D} (without rim) and \hyperref[model:E]{E} (with rim)} for the inner and outer race of an unloaded contact of a 6205-C3 bearing with $\varepsilon_\mathrm{Oil}=2.2$.}
		\label{fig:Semi3DRaceRim}
	\end{figure}

The semi-analytical method allows to consider not only the groove geometry, \rev{method \hyperref[model:D]{D},} but also the rim area aside of the groove\rev{, method \hyperref[model:D]{D}}. A comparison of the calculated capacitance of unloaded rolling elements over a varying lubrication gap height is shown in Fig. \ref{fig:Semi3DRaceRim}. For a bigger lubrication gap, the influence of the rim area increases. The rolling elements with a high lubrication gap contribute rather little to the total capacitance, thus, the rim area may be neglected for typical applications but should be considered for high precision calculations. Therefore, it is considered in this work for better comparability with the 3D FE model \hyperref[model:H]{H} which by nature considers this area.

\subsection{\rev{3D FE simulation results}}
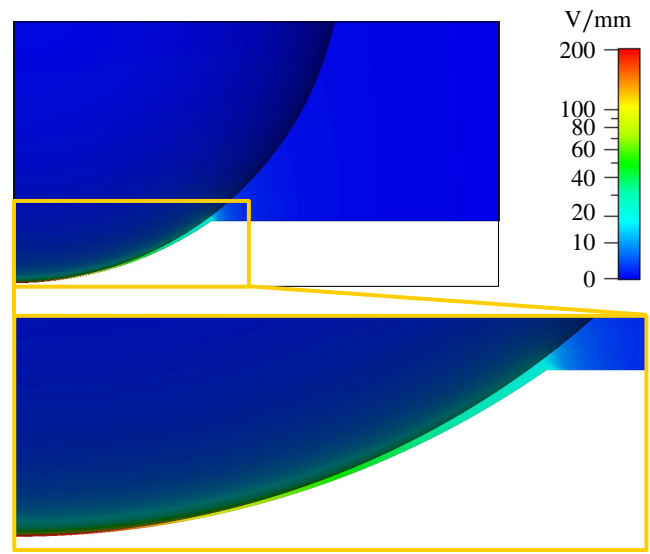
\begin{figure}[pos=htb]
		\centering
		\begin{tikzpicture}
			\node [above right, inner sep=0] (image) at (0,0) {\includegraphics[width=\linewidth]{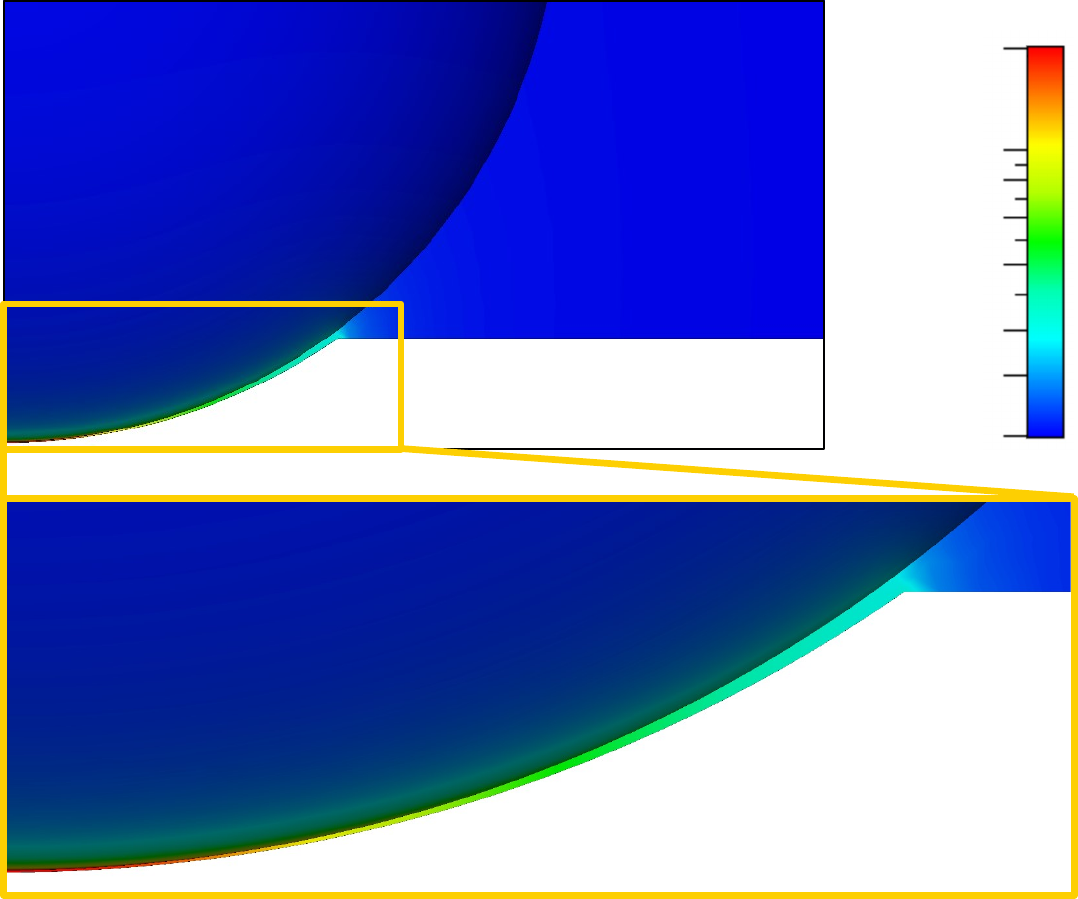}};
			\begin{scope}[
				x={($0.1*(image.south east)$)},
				y={($0.1*(image.north west)$)}]
				%\draw[lightgray,step=1] (image.south west) grid (image.north east);
				
				\draw (9.2,10   ) node{$\mathrm{V}/\mathrm{mm}$};
				\draw (9.3,5.15) node[anchor=east]{$0$};
				\draw (9.3,5.85) node[anchor=east]{$10$};
				\draw (9.3,6.4  ) node[anchor=east]{$20$};
				\draw (9.3,7.05) node[anchor=east]{$40$};
				\draw (9.3,7.6  ) node[anchor=east]{$60$};
				\draw (9.3,8     ) node[anchor=east]{$80$};
				\draw (9.3,8.35) node[anchor=east]{$100$};
				\draw (9.3,9.45) node[anchor=east]{$200$};
			\end{scope}
		\end{tikzpicture}
		\caption{Electric field strength by 3D FE simulation in section plane I for the outer ring for the bearing 6205-C3 and an lubrication gap of $5 \mathrm{\mu m}$.}
		\label{fig:3DnumericFieldStrength}
	\end{figure}

The resulting absolute values of the electric field strength as shown in Fig. \ref{fig:3DnumericFieldStrength} for section plane I indicate a very strong concentration of the electric field in the lubrication gap. At an exemplary lubrication gap of $5 \mathrm{\mu m}$ the field strength drops from the center to the edge of the groove by a factor of more than $10$. As expected, the field strength outside the groove drops rapidly due to the increasing distance between the race and the ball. Accordingly, the influence of boundary condition selection is low.

	\begin{figure}[pos=htb]
		\centering
		\begin{tikzpicture}
			\node [above right, inner sep=0] (image) at (0,0) {\includegraphics[width=\linewidth]{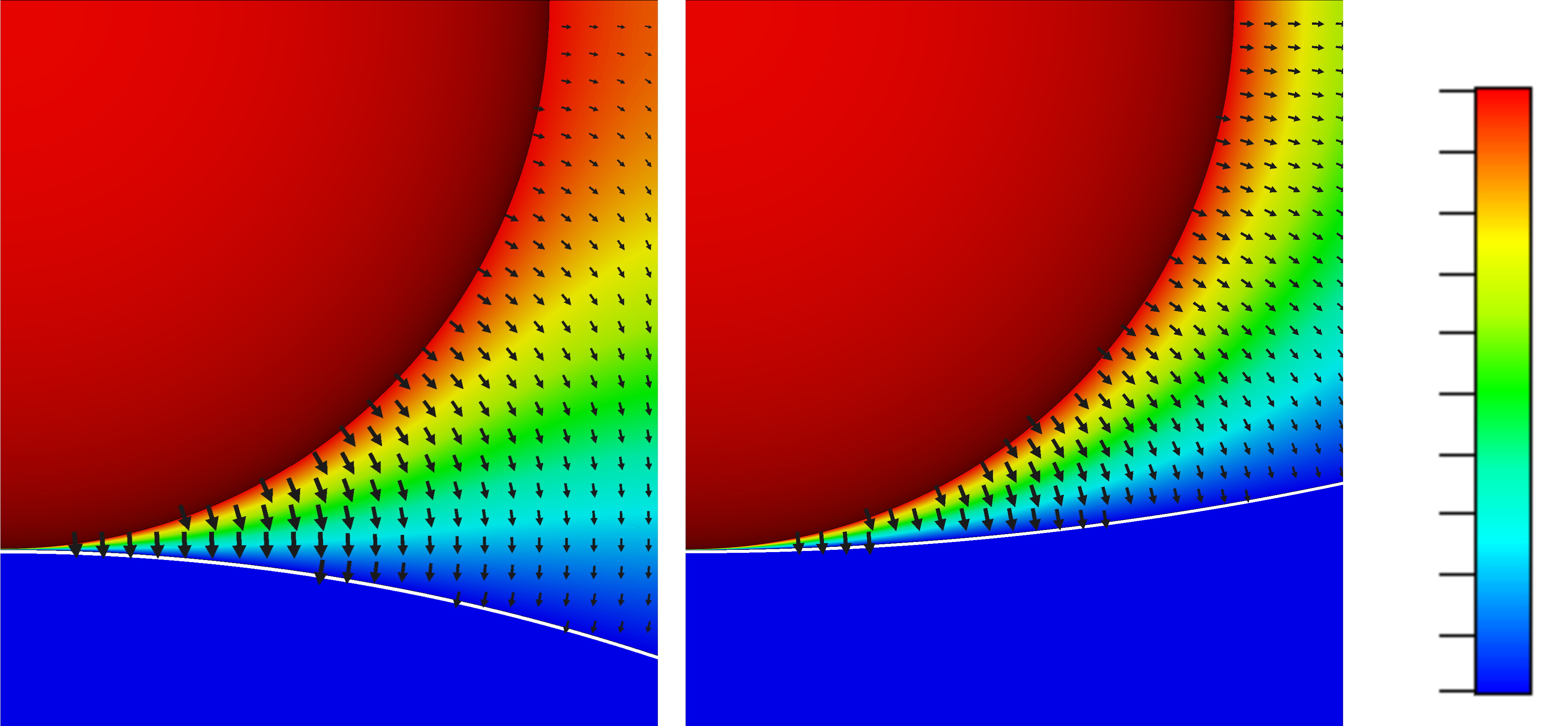}};
			\begin{scope}[
				x={($0.1*(image.south east)$)},
				y={($0.1*(image.north west)$)}]
				%\draw[lightgray,step=1] (image.south west) grid (image.north east);
				
				\draw (9.2,9.4 ) node{$\mathrm{V}$};
				\draw (8.5,0.5 ) node[anchor=west]{$0$};
				\draw (8.5,1.33) node[anchor=west]{$0.1$};
				\draw (8.5,2.16) node[anchor=west]{$0.2$};
				\draw (8.5,2.99) node[anchor=west]{$0.3$};
				\draw (8.5,3.82) node[anchor=west]{$0.4$};
				\draw (8.5,4.65) node[anchor=west]{$0.5$};
				\draw (8.5,5.48) node[anchor=west]{$0.6$};
				\draw (8.5,6.31) node[anchor=west]{$0.7$};
				\draw (8.5,7.14) node[anchor=west]{$0.8$};
				\draw (8.5,7.97) node[anchor=west]{$0.9$};
				\draw (8.5,8.8 ) node[anchor=west]{$1$};
			\end{scope}
		\end{tikzpicture}
		\caption{Potential fields of the 3D FE simulation in section plane II for the outer (left) and inner (right) ring for the bearing 6205-C3 and a lubrication gap of $5 \mathrm{\mu m}$.}
		\label{fig:3DnumericResults}
	\end{figure}
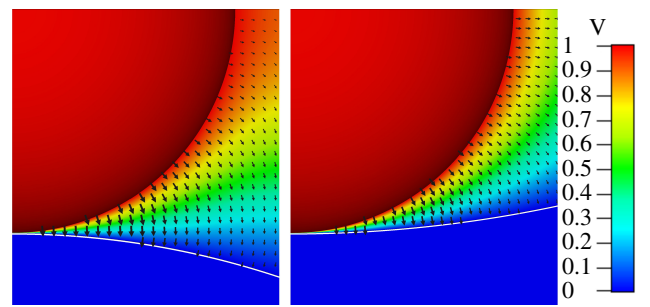

Fig. \ref{fig:3DnumericResults} displays the direction of the electric field as well as the potential field in section plane II for the inner and outer ring. At the outer ring, the equipotential planes are denser resulting in a higher electric field and therefore a higher capacitance.
	
\subsection{Comparison of 3D results}
\label{seq:comparison3D}
\begin{figure}[pos=htb]
		\centering
		\begin{tikzpicture}
			\node [above right, inner sep=0] (image) at (0,0) {\includegraphics[width=\linewidth]{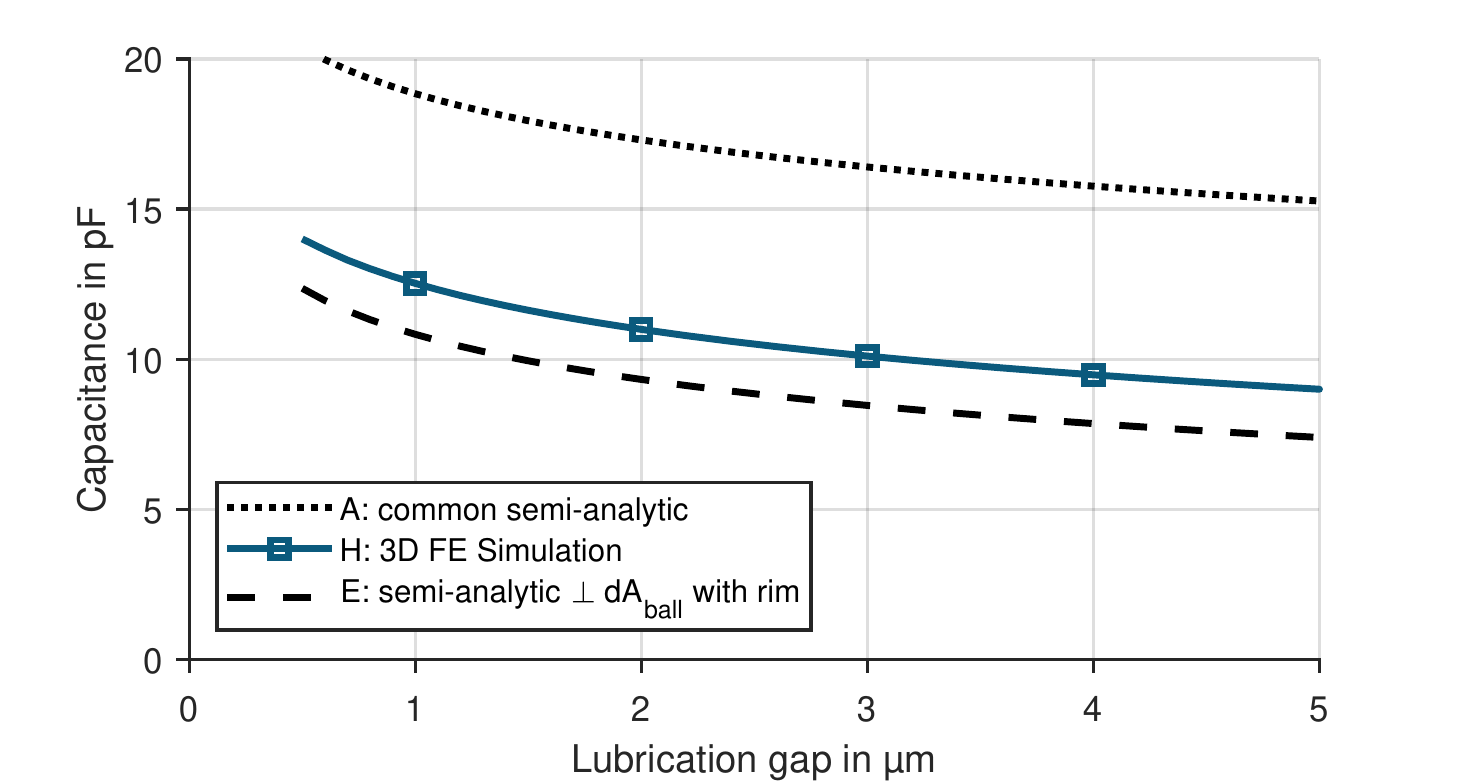}};
			\begin{scope}[
				x={($0.1*(image.south east)$)},
				y={($0.1*(image.north west)$)}]
				%\draw[lightgray,step=1] (image.south west) grid (image.north east);
			\end{scope}
		\end{tikzpicture}
		\caption{Comparison of 3D models \rev{\hyperref[model:A]{A}, \hyperref[model:E]{E} and \hyperref[model:H]{H}} for the inner race of a 6205-C3 radial deep-groove ball bearing with $\varepsilon_{\mathrm{Oil}}=2.2$.}
		\label{fig:3Dcomparison}
	\end{figure}

\rev{The common semi-analytical model \hyperref[model:A]{A} and the semi-analytic method \hyperref[model:E]{E} ($\perp\mathrm{d}A_{\mathrm{ball}}$ with rim) are compared with the 3D FE model \hyperref[model:H]{H} in Fig. \ref{fig:3Dcomparison}. The method using the effective radii overestimates the 3D FE simulation results throughout, following the trend observed in 2D, Fig. \ref{fig:2Dcomparison}. This is due to the neglect of the groove width in the method of effective radii. Method \hyperref[model:E]{E}, using the true geometry, takes the groove width into account as integration limits. Thus, a much smaller deviation from the 3D FE model \hyperref[model:H]{H} can be achieved, which also corresponds to Fig. \ref{fig:2Dcomparison}. The remaining deviations can be explained by higher deviations in section plane II as compared to section plane I, Fig. \ref{fig:2Dcomparison}.} On top of that, the simulation also takes stray capacitances \rev{beyond the rim area} into account, which are neglected in the semi-analytic method.

\section{Conclusions}

\begin{revision}
    The capacitance of a rolling contact can be calculated in various ways. The most used semi-analytic approximation considers the lubrication gap as a parallel connection of infinitesimal plate capacitors, from which the full capacitance value follows by integration. Commonly, the geometry is simplified to a sphere on plane contact, where the effective radii of the sphere in both section planes are found as half of the harmonic mean of the radii of the rolling element and the raceway (model \hyperref[model:A]{A}). The assumptions of this approach have been investigated and quantified. For instance, the groove width is not considered, thus, deviations are found in 2D from a newly derived analytic solution (model \hyperref[model:F]{F}) and a 2D FE simulation (model \hyperref[model:G]{G}), which are in good accordance to each other. A comparison with a 3D FE simulation (model \hyperref[model:H]{H}) shows the same results.
    
    Four additional semi-analytic models have been developed using the true geometry: assuming vertically oriented electric field lines (model \hyperref[model:B]{B}) or assuming electric field lines perpendicular to the rolling element (models \hyperref[model:C]{C}-\hyperref[model:E]{E}), deriving infinitesimal area elements from the raceway surface (model \hyperref[model:C]{C}) or from the rolling element surface (models \hyperref[model:D]{D} and \hyperref[model:E]{E}) while considering only the groove (models \hyperref[model:C]{C} and \hyperref[model:D]{D}) or including the rim aside of the groove (model \hyperref[model:E]{E}). Especially model \hyperref[model:E]{E} is recommended as it allows to account for the rim area aside of the groove. Thus, it is almost as powerful as the 3D FE model \hyperref[model:H]{H} and delivers a viable advancement to the state of the art models.
\end{revision}
	
%\pagebreak

\appendix 
\section{\rev{Summary of capacitance models used}} \label{seq:appendix}

\begin{enumerate}[A:]
    \item semi-analytic common \dotfill Section \ref{seq:bearingCapacitance}
    \item semi-analytic $\|$ \dotfill Section \ref{model:B}
    \item semi-analytic $\perp\mathrm{d}A_{\mathrm{race}}$   \dotfill Section \ref{model:C}
	\item semi-analytic $\perp\mathrm{d}A_{\mathrm{ball}}$   \dotfill Section \ref{model:D}
	\item semi-analytic $\perp\mathrm{d}A_{\mathrm{ball}}$ with rim   \dotfill Section \ref{model:E}
	\item analytic   \dotfill Section \ref{model:F}
	\item 2D FE simulation   \dotfill Section \ref{model:G}
	\item 3D FE simulation   \dotfill Section \ref{model:H}
\end{enumerate}

\printcredits

\section*{Funding Acknowledgement}
This work was funded by the Deutsche Forschungsgemeinschaft (DFG, German Research Foundation) under the project number 467849890.

\bibliographystyle{model3-num-names}

\bibliography{cas-refs}

\end{document}